\documentclass[10pt,a4paper]{article}
\usepackage[utf8]{inputenc}
\usepackage{a4wide}
\usepackage{graphicx}
\usepackage{hyperref}
\usepackage{color}

\usepackage{amssymb}
\usepackage{amsmath}
\usepackage{verbatim} 

\begin{document}

\begin{titlepage}
\begin{flushright}
\end{flushright}
\vfill
\begin{center}
{\Large\bf 	New analytic continuations for the Appell $F_4$ series}
\end{center}

\begin{center}
{\Large\bf from quadratic transformations of the Gauss $_{2}F_1$ function}

\vfill
{\bf B. Ananthanarayan$^a$, Samuel Friot~$^{b,c}$, Shayan Ghosh~$^d$ and Anthony Hurier$^{a,b,e}$\footnote{A. Hurier was awarded the Raman-Charpak fellowship by CEFIPRA for part of his work on the present subject.}}\\[1cm]
{$^a$ Centre for High Energy Physics, Indian Institute of Science, \\
Bangalore-560012, Karnataka, India}\\[0.5cm]
{$^b$ Universit\'e Paris-Saclay, CNRS/IN2P3, IJCLab, 91405 Orsay, France } \\[0.5cm]
{$^c$ Univ Lyon, Univ Claude Bernard Lyon 1, CNRS/IN2P3, IP2I Lyon,\\
 UMR 5822, F-69622, Villeurbanne, France}\\[0.5cm]
{$^d$ Helmholtz-Institut f\"ur Strahlen- und Kernphysik \& Bethe Center for Theoretical Physics, Universit\"at Bonn, D-53115 Bonn, Germany} \\[0.5cm]
{$^e$ Sorbonne Universit\'e, Facult\'e des Sciences et Ing\'enierie,\\
 75005, Paris, France}\\
\end{center}
\vfill
\begin{abstract}
We present new analytic continuation formulas for the Appell $F_4(a,b;c,d;x,y)$ double hypergeometric series where $d=a-b+1$, which allows quadratic transformations of the Gauss ${}_2F_1$ hypergeometric function to be used in the intermediate steps of the derivation. Such formulas are of relevance to loop calculations of quantum field theory where they can been used, for instance, to obtain new series representations of the two-loop massive sunset Feynman diagram. The analytic continuation procedure introduced in this paper is also sufficiently general so as to find uses elsewhere. 
\end{abstract}
\vfill
\vfill
\end{titlepage}

\section{Introduction}

While evaluating Feynman integrals analytically in perturbative calculations of quantum field theory, multiple hypergeometric series and their generalizations are natural objects appearing when one uses computational techniques like for instance the Mellin-Barnes (MB) representation method. Large classes of these series have been well-studied in the mathematical literature for a long time, allowing one to use some of their interesting properties like reduction formulas, convergence theorems or analytic continuation results, etc. (see \cite{Srivastava} for an excellent introduction). Although beyond triple hypergeometric series the mathematical literature is less rich, due to the growth of complexity of these objects, new results still appear regularly in this classical branch of mathematics, possibly leading to applications in the field of investigations related to Feynman integrals.
One may note, however, that things can also happen the other way around, since works motivated by Feynman integrals studies can also enrich the multiple hypergeometric series theory, either at a fundamental level or at a more practical one, see for instance \cite{Abreu:2019wzk, Brown:2019jng, Bytev:2020zhg,Shpot1,Shpot:2007bz}. 

The results presented in this paper are in line with this latter approach, since they were derived for the need of obtaining new triple series representations \cite{Ananthanarayan:2019icl}, relevant for phenomenological purposes, of the two loop massive sunset diagram, by performing analytic continuations of well-known expressions of the latter given in \cite{Berends:1993ee} in terms of Lauricella $F_C$ triple series.  
Indeed, if the numerical evaluation of multiple hypergeometric series from their partial sums is easy to implement on a computer and often enough precise, which renders them attractive in view of physical applications, the expressions of Feynman integrals in terms of these series obviously give only a restricted mathematical description, valid in their convergence domains. Furthermore, once the proper regions of convergence of the involved series have been found (which, already at the level of triple series, can be a science in itself \cite{Srivastava}), if one is fortunate enough the physical values of their expansion parameters such as combinations of masses, kinematical invariants, etc. belong to their intersection. It is otherwise necessary to proceed to analytic continuations of the series, which was the case in \cite{Ananthanarayan:2019icl}. It is possible to extract some of these analytic continuations, in general the most trivial ones, from the MB method mentioned above. This has been for instance systematized in \cite{Friot:2011ic} for the case of twofold MB integrals. However, the examples considered therein show that, even at the twofold level, in general some region in the parameter space remains, that cannot be reached by any of the different series representations deduced from a standard residue calculation of the MB representation  (see however \cite{Ananthanarayan:2020acj} for a twofold example where such a region does not appear). This region corresponds to other non-trivial analytic continuations and, to access them, alternative techniques have to be used. This is precisely the subject of the present work, where we consider a particular case of the double hypergeometric  Appell $F_4$ series for which, using quadratic transformations of the Gauss hypergeometric ${}_2F_1$ function in the intermediate steps of their derivations, we will present new analytic continuation formulas that cannot be obtained from a direct residue computation of the Mellin-Barnes representation of the corresponding Appell $F_4$ function.

Appell functions have been defined by P. Appell in 1880 (\cite{Kampe}) and are of four different types: $F_i$ with $i=1,...,4$. The last of them, $F_4(a,b;c,d;x,y)$, which is also the most complicated to handle, has four parameters $a,b,c$ and $d$, and two variables $x$ and $y$. Appell functions appear in several instances in quantum field theory. To mention just a few cases that concern $F_4$, the latter is for instance the main ingredient in the mathematical expression of the two-loop vacuum sunset diagram which is the building block of a computational method presented in  \cite{DT}, whose aim is to obtain power series in the external momentum of two-loop self-energy Feynman diagrams with arbitrary masses. The Appell $F_4$ function also appears in the expression of the one-loop triangle diagrams considered in \cite{Alkofer:2008dt} for the study of the infrared behaviour of three-point functions in Landau gauge Yang-Mills theory. The authors of \cite{Alkofer:2008dt} could not use the common series representation of $F_4$ for numerical purposes because they needed it in the Euclidean momentum region which lies outside of the boundary of its known convergence region. Therefore, they have been forced to use analytic continuations of the $F_4$ series, taken from \cite{Exton}.

 It is interesting to note that the latter are also useful on the experimental side, where $F_4$ appears in the completely different context of the physics of detectors. Depending on the distance of a disk source to a coaxial parallel disk detector and of their radii, the computation of the corresponding solid angle, whose closed expression is in terms of $F_4$, will necessitate the use of its common series representation or its analytic continuations \cite{FriotRuby}. As a last example, which we recall to be the main motivation for the present study, let us cite the occurence of the $F_4$ series in the analytic expressions of the two-loop sunset diagram with four mass scales already mentioned above \cite{Berends:1993ee} since Lauricella $F_C$ series can be seen as infinite sums of $F_4$ series. The $F_C$ series can then be analytically continued using some of the new analytic continuation formulas of $F_4$ that we give in the present paper. It may be noted that the analytic continuation formulas given in \cite{Exton} are not well suited for the particular case of the sunset diagram with four mass scales studied in \cite{Ananthanarayan:2019icl}, thereby providing a further justification for the present work.

Let us now briefly summarize the content of the paper. In Section \ref{AppellF4def}, we recall basic facts about the Appell $F_4$ function. In Section \ref{AppellF4AC} the derivation of new analytic continuations formulas is detailed and in section \ref{other} we present some alternative results. Section \ref{concl} contains our conclusions and an appendix follows, where a numerical illustration of some subtelties that one must take care of when using the quadratic transformations is presented.

\section{The Appell $F_4$ function \label{AppellF4def}}
In this section we present some well-known facts about the Appell $F_4$ function. 

\noindent Its series representation in the vicinity of $(x,y)=(0,0)$ is
\begin{align}
F_{4}(a,b;c,d;x,y)=\sum_{m,n=0}^{\infty}\frac{(a)_{m+n}(b)_{m+n}}{(c)_m(d)_n}\frac{x^m}{m!}\frac{y^n}{n!}\ .
\label{F4}
\end{align}
The Appell $F_4$ series shown in Eq.(\ref{F4}) is one of the four Appell series which are double series generalisations of the Gauss hypergeometric series
\begin{align}
{}_{2}F_{1}(a,b;c;x)=\sum_{k=0}^{\infty}\frac{(a)_k(b)_k}{(c)_k}\frac{x^k}{k!}\ ,
\end{align}
where $(a)_k=\frac{\Gamma(a+k)}{\Gamma(a)}$ is the Pochhammer symbol. 

\begin{figure}[hbtp]
\centering
\includegraphics[scale=0.5]{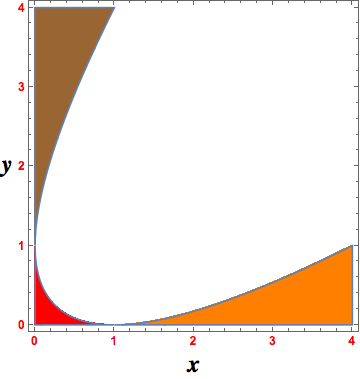}
\caption{Convergence regions $R$ (red) and $R'$ (brown) for real positive values of $x$ and $y$. For the orange region, see the text.}
\label{Fig1}
\end{figure}

The convergence region $R$ of the $F_4$ series in Eq.(\ref{F4}) is $R\doteq \sqrt{\vert x\vert }+\sqrt{\vert y\vert }<1$. For real positive values of $x$ and $y$, this is a quite restricted area in the $(x,y)$-plane, see the red region in Fig. \ref{Fig1}, but simple analytic continuations of Eq.(\ref{F4}) allow for some series representations which are valid in the vicinity of $(0,\infty)$ and $(\infty,0)$. 
Near $(0,\infty)$, for instance, we have \cite{Kampe}
\begin{align}
	F_4(a,b;&c,d;x,y) = \nonumber \\
	&\frac{\Gamma(d)\Gamma(b-a)}{\Gamma(d-a)\Gamma(b)} (-y)^{-a}
	 F_4\left(a,a-d+1;c,a-b+1;\frac{x}{y},\frac{1}{y}\right) \nonumber \\
	 +&\frac{\Gamma(d)\Gamma(a-b)}{\Gamma(d-b)\Gamma(a)} (-y)^{-b}
	 F_4\left(b,b-d+1;c,b-a+1;\frac{x}{y},\frac{1}{y}\right)
	 \label{F4_AC1}
\end{align}
where the series in the r.h.s. of Eq.(\ref{F4_AC1}) converge in the region $R'\doteq \sqrt{\left\vert\frac{x}{y}\right\vert }+\sqrt{\left\vert\frac{1}{y}\right\vert }<1$ shown in brown in Fig. \ref{Fig1}. 

By the exchange of $x$ and $y$ as well as $c$ and $d$ which leaves the l.h.s of Eq.(\ref{F4_AC1}) unchanged, it is easy to obtain another analytic continuation formula which converges in the symmetric orange region of Fig. \ref{Fig1}. These analytic continuation formulas as well as Eq.(\ref{F4}) are the only series representations of the Appell $F_4$ function that may be obtained by a direct residues computation of its Mellin-Barnes representation \cite{Kampe}
\begin{align}
	F_4(a,b;&c,d;x,y) = \frac{\Gamma(c)\Gamma(d)}{\Gamma(a)\Gamma(b)}\frac{1}{(2i\pi)^2}\int_{L}\int_{L'}ds\ dt(-x)^{s}(-y)^{t}\frac{\Gamma(-s)\Gamma(-t)\Gamma(a+s+t)\Gamma(b+s+t)}{\Gamma(c+s)\Gamma(d+t)}
	 \label{F4_MB}
\end{align}
where the contours $L$ and $L'$ are defined in the usual way \cite{Kampe}.

Alternatively, the analytic continuation formulas presented above may be derived from a rewriting of the Appell $F_4$ series in Eq. (\ref{F4}) as a sum of Gauss ${}_2F_1$ hypergeometric series on which one can apply a well-known analytic continuation.

We can see that in Fig. \ref{Fig1} there remains a large part of the $(x,y)$-plane (in what follows we refer to it as the ``white region") which is not reached by any of the three standard series representations of the Appell $F_4$ function that we have listed above. As already mentioned in the introduction, the white region of Fig. \ref{Fig1} exactly corresponds to the Euclidean momentum region studied in \cite{Alkofer:2008dt}  where closed expressions of the massless one-loop triangle diagrams are given in terms of $F_4$ with variables $x=\tfrac{p_1^2}{p_3^2}$, $y=\tfrac{p_2^2}{p_3^2}$. Therefore, it was necessary for the authors of \cite{Alkofer:2008dt} to use other analytic continuations of $F_4$, namely the non-trivial ones derived in \cite{Exton}, since the latter are valid in this particular region.

In the next section, we show that some alternative analytic continuation formulas can be obtained. They are restricted to the case where $d=a-b+1$ but, although less general than those of \cite{Exton}, they are better suited for an application to the case of the sunset diagram with four mass scales, in order to obtain series representations of the latter valid for chiral perturbation theory (ChPT) or for the minimal supersymmetric standard model (MSSM) \cite{Ananthanarayan:2019icl}.

\section{New analytic continuation formulas for Appell $F_4$ series\label{AppellF4AC}}
It is not possible to find analytic continuation formulas,  different from those presented in the previous section, by standard residues computations of the Mellin-Barnes representation of the Appell $F_4$ function as it stands in Eq.(\ref{F4_MB}). However, the second strategy mentioned in Section \ref{AppellF4def} where $F_4$ is rewritten as a sum of ${}_2F_1$ may be of help to accomplish this task, and this is indeed the case as we will see now in detail. In the case of the sunset diagram with four mass scales, the particular combination $d=a-b+1$ (or $c=a-b+1$) appears \cite{Ananthanarayan:2019icl}. Since the sunset was the main motivation for the present work, we will stay in this specific case. Indeed, as a consequence of this particular pattern, the use of quadratic transformation formulas of ${}_2F_1$, leading to new analytic continuations of $F_4$ well suited for the sunset diagram, becomes possible.

In light of the comments above, let us begin by rewriting Eq.(\ref{F4}) as 
\begin{align}
F_{4}(a,b;c,a-b+1;x,y) &= \sum\limits_{m=0}^{\infty} \frac{(a)_{m} (b)_{m}}{(c)_{m}} \frac{x^m}{m!} {}_{2}F_{1}(a+m,b+m;a-b+1;y).
\label{F42F1}
\end{align}
From the pattern of the parameters of the Gauss ${}_2F_1$ hypergeometric series in the r.h.s. of Eq.(\ref{F42F1}), one can apply the following quadratic transformation formula \cite{A&S}
\begin{align}
{}_{2}F_{1}(\alpha,\beta;\alpha-\beta+1;y) &= (1-y)^{-\alpha} {}_{2}F_{1}\left(\frac{\alpha}{2},  \frac{\alpha}{2}-\beta+\frac{1}{2}; \alpha-\beta+1;-\frac{4 y}{(1-y)^2}\right),
\label{quadratic1}
\end{align}
which is valid for $|y|<1$.

Inserting Eq.(\ref{quadratic1}) in Eq.(\ref{F42F1}) we get, for $|y|<1$,
\begin{align}
F_{4}(a,b;c,a-b+1;x,y) &= \sum\limits_{m=0}^{\infty} \frac{(a)_{m} (b)_{m}}{(c)_{m}} \frac{x^m}{m!} (1-y)^{-a-m}\nonumber \\ &\times {}_{2}F_{1}\left(\frac{a+m}{2}, \frac{a-m+1}{2}-b; a-b+1;\frac{- 4 y}{(1-y)^2}\right).
\end{align}
It is now possible to apply on the r.h.s. the following analytic continuation of ${}_2F_1$ \cite{A&S}:
\begin{align}
{}_{2}F_{1}(\alpha,\beta;\gamma;x) &= (1-x)^{-\alpha} \frac{\Gamma(\gamma) \Gamma(\beta-\alpha)}{\Gamma(\beta) \Gamma(\gamma-\alpha)} {}_{2}F_{1}\left(\alpha;\gamma-\beta,\alpha-\beta+1;\frac{1}{1-x}\right)\nonumber \\ &+ (1-x)^{-\beta} \frac{\Gamma(\gamma) \Gamma(\alpha-\beta)}{\Gamma(\alpha) \Gamma(\gamma-\beta)} {}_{2}F_{1}\left(\beta,\gamma-\alpha;\beta-\alpha+1;\frac{1}{1-x}\right),
\label{2F1AC}
\end{align}
which, in principle, is valid only for $\vert \arg(1-x)\vert<\pi$, the latter condition being satisfied here. Note however that the usual convention to define values of ${}_2F_1(\alpha, \beta;\gamma,x)$ on its cut (\textit{i.e} for $x>1$) from the limits of values in the lower half complex $x$-plane (the Counter Clockwise Continuity (CCC)) allows to discard the $\vert \arg(1-x)\vert<\pi$ constraint and Eq.(\ref{2F1AC}) becomes valid for all complex values of $x$ \cite{BS}.

\begin{figure}[hbtp]
\centering
\includegraphics[scale=0.5]{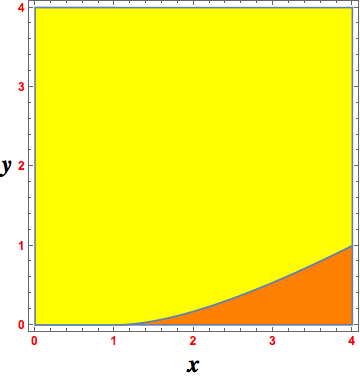}
\caption{Convergence region $R_0$, in yellow.}
\label{Fig2}
\end{figure}

One then obtains, after rewriting the two ${}_2F_1$ as series and using the duplication and generalized reflection formulas for the Euler $\Gamma$ function,
\begin{align}
F_{4}(a,b;c,a-b+1;x;y)=(1-y)^{-a}\left(\frac{(1+y)^2}{(1-y)^2}\right)^{-\frac{a}{2}}\frac{2^{a-2b}}{\sqrt{\pi}}\frac{\Gamma(a-b+1)\Gamma\left(-b+\frac{1}{2}\right)\Gamma\left(\frac{1}{2}+b\right)\Gamma(c)}{\Gamma(a-2b+1)\Gamma(2b-a)\Gamma(a)\Gamma(b)}\nonumber\\
\times \sum_{m,n=0}^\infty\left[\frac{x}{2(1-y)}\left(\frac{(1+y)^2}{(1-y)^2}\right)^{-\frac{1}{2}}\right]^m\left[\frac{(1-y)^2}{4(1+y)^2}\right]^n\frac{1}{m!n!}\frac{\Gamma(b+m)\Gamma(m-a+2b)\Gamma(a+m+2n)}{\Gamma(c+m)\Gamma(b+\frac{1}{2}+m+n)}\nonumber\\
+(1-y)^{-a}\left(\frac{(1+y)^2}{(1-y)^2}\right)^{-\frac{a}{2}+b-\frac{1}{2}}\frac{2^{a-1}}{\sqrt{\pi}}\frac{\Gamma(a-b+1)\Gamma\left(-b+\frac{3}{2}\right)\Gamma\left(-\frac{1}{2}+b\right)\Gamma(c)}{\Gamma(a-2b+1)\Gamma(2b-a)\Gamma(a)\Gamma(b)}\nonumber\\
\times \sum_{m,n=0}^\infty\left[\frac{2x}{1-y}\left(\frac{(1+y)^2}{(1-y)^2}\right)^{\frac{1}{2}}\right]^m\left[\frac{(1-y)^2}{4(1+y)^2}\right]^n\frac{1}{m!n!}\frac{\Gamma(b+m)\Gamma(a-2b+1+2n-m)\Gamma(m-a+2b)}{\Gamma(c+m)\Gamma(-b+\frac{3}{2}-m+n)}.
\label{firstAC}
\end{align}
For real values of $x$ and $y$ one could give more compact expressions for some of the combinations of $x$ and $y$ that appear in the r.h.s. of Eq.(\ref{firstAC}) but we keep the latter as they are written in order that the equation stays valid in the complex case.
We now have to find the convergence region of each of the two double series of the new series representation of $F_4$ given in Eq.(\ref{firstAC}), to see if this formula is of interest. Let us consider the first series.
We note that it has the same convergence behaviour as the series:
\begin{align}
	\sim \sum_{m,n=0}^\infty \frac{X^m}{m!} \frac{Y^n}{n!}  \frac{\Gamma(A+m+2n) \Gamma(B+m) \Gamma(C+m)}{\Gamma(D+m+n) \Gamma(E+m)}
	\label{equivSeries1}
\end{align}
with
\begin{align}
	 X = \frac{x}{2(1-y)}\left(\frac{(1+y)^2}{(1-y)^2}\right)^{-\frac{1}{2}} 
\end{align}	
and
\begin{align}
	 Y = \frac{(1-y)^2}{4(1+y)^2}.
\end{align}
This is due to the fact that the convergence properties of multiple gaussian hypergeometric series are independent of their parameters (exceptional parameter values being excluded) \cite{Srivastava}. Therefore, one can go one step further and choose $C=E$ and, by cancellation of the two corresponding $\Gamma$ functions in Eq.(\ref{equivSeries1}), we find that our series behaves in fact like the Horn series $H_3(A,B;C;X,Y)$ whose properties are well-known \cite{Srivastava}, its convergence region being:
\begin{align}
	\vert Y\vert <\frac{1}{4} \wedge \vert X\vert <\frac{1}{2}+\frac{1}{2}\sqrt{1-4\vert Y\vert}.
\end{align}
Then, the first series of our analytic continuation formula converges in the region
\begin{align}
	R_0\doteq\left\vert \frac{(1-y)^2}{4(1+y)^2}\right\vert <\frac{1}{4} \wedge \left\vert \frac{x}{2(1-y)}\left(\frac{(1+y)^2}{(1-y)^2}\right)^{-\frac{1}{2}}\right\vert <\frac{1}{2}+\frac{1}{2}\sqrt{1-4\left\vert \frac{(1-y)^2}{4(1+y)^2}\right\vert}
	\label{rzero}
\end{align}
shown in yellow in Fig. \ref{Fig2}. We can see that it includes the full desired region, i.e. the one which is not accessed in Fig. \ref{Fig1} (the "white" region). This is a very good beginning.
We now consider the second series in Eq.(\ref{firstAC}). Performing a similar analysis, we find that it behaves like the Horn series $H_6(A,B,C;-X,-Y)$ (see \cite{Srivastava} for details about this series) with $X=\frac{2x}{1-y}\left(\frac{(1+y)^2}{(1-y)^2}\right)^{\frac{1}{2}}$ and $Y=\frac{(1-y)^2}{4(1+y)^2}$, giving a convergence region which, in fact, is included in $R\cup R'$ and, therefore, not interesting: it is necessary to go on the analytic continuation process for this series.

In order to achieve this goal, we will now consider the sum over $m$ of the latter. Indeed, until now we have performed manipulations only on the sum over $n$. The sum over $m$ in the second series of Eq.(\ref{firstAC}) may be written as
\begin{align}
&\sum_{m=0}^\infty \left[\frac{2x}{1-y}\left(\frac{(1+y)^2}{(1-y)^2}\right)^{\frac{1}{2}}\right]^m\frac{1}{m!}\frac{\Gamma(b+m)\Gamma(a-2b+1+2n-m)\Gamma(m-a+2b)}{\Gamma(c+m)\Gamma(-b+\frac{3}{2}-m+n)}\nonumber\\
&=\frac{\Gamma(a-2b+1+2n)\Gamma(b)\Gamma(2b-a)}{\Gamma\left(\frac{3}{2}-b+n\right)\Gamma(c)} {}_3F_2\left(b,-\frac{1}{2}+b-n,2b-a;c,2b-a-2n;\frac{2x}{1-y}\left(\frac{(1+y)^2}{(1-y)^2}\right)^{\frac{1}{2}}\right).
\label{3F2avantMB}
\end{align}
Now we use the Mellin Barnes representation of the ${}_3F_2$ generalized hypergeometric function
\begin{equation}
{}_3F_2\left(\alpha,\beta,\gamma;\delta,\epsilon;z\right)=\frac{\Gamma(\delta)\Gamma(\epsilon)}{\Gamma(\alpha)\Gamma(\beta)\Gamma(\gamma)}\frac{1}{2 i \pi}\int_{L}  ds (-z)^{-s}\frac{\Gamma(s)\Gamma(\alpha-s)\Gamma(\beta-s)\Gamma(\gamma-s)}{\Gamma(\delta-s)\Gamma(\epsilon-s)}
\label{3F2MB}
\end{equation}
where the $L$-contour separates the poles of $\Gamma(s)$ from the poles of the other Gamma functions in the numerator. By closing the contour to the right, we get
\begin{align}
&\frac{1}{2 i \pi}\int_{L}  ds \left(-\frac{2x}{1-y}\left(\frac{(1+y)^2}{(1-y)^2}\right)^{\frac{1}{2}}\right)^{-s}\frac{\Gamma(s)\Gamma(b-s)\Gamma(-\frac{1}{2}+b-n-s)\Gamma(2b-a-s)}{\Gamma(c-s)\Gamma(2b-a-2n-s)} \nonumber\\
=&\left(-\frac{2x}{1-y}\left(\frac{(1+y)^2}{(1-y)^2}\right)^{\frac{1}{2}}\right)^{-b}\sum_{m=0}^\infty\left(\frac{1-y}{2x}\left(\frac{(1-y)^2}{(1+y)^2}\right)^{\frac{1}{2}}\right)^{m}\frac{\Gamma(b+m)\Gamma(-\frac{1}{2}-n-m)\Gamma(b-a-m)}
{m!\ \Gamma(c-b-m)\Gamma(b-a-2n-m)} \nonumber \\
+&\left(-\frac{2x}{1-y}\left(\frac{(1+y)^2}{(1-y)^2}\right)^{\frac{1}{2}}\right)^{\frac{1}{2}-b+n}\sum_{m=0}^\infty\left(\frac{1-y}{2x}\left(\frac{(1-y)^2}{(1+y)^2}\right)^{\frac{1}{2}}\right)^{m} \nonumber\\
&\hspace{4.5cm}\times\frac{\Gamma(b-\frac{1}{2}-n+m)\Gamma(\frac{1}{2}+n-m)\Gamma(b-a+\frac{1}{2}+n-m)}{m!\ \Gamma(c-b+\frac{1}{2}+n-m)\Gamma(b-a+\frac{1}{2}-m-n)}\nonumber\\
+&\left(-\frac{2x}{1-y}\left(\frac{(1+y)^2}{(1-y)^2}\right)^{\frac{1}{2}}\right)^{-2b+a}\sum_{m=0}^\infty\left(\frac{y-1}{2x}\left(\frac{(1-y)^2}{(1+y)^2}\right)^{\frac{1}{2}}\right)^{m} \nonumber\\
&\hspace{4.5cm}\times\frac{\Gamma(m+2b-a)\Gamma(a-b-m)\Gamma\left(a-b-\frac{1}{2}-m-n\right)\Gamma(1+m+2n)}{m!\Gamma(c-2b+a-m)\Gamma(-2n)\Gamma(1+2n)}.
\label{MBexp}
\end{align}

Note that while the first two sums in the r.h.s. converge for $\left\vert\frac{1-y}{2x}\left(\frac{(1-y)^2}{(1+y)^2}\right)^{\frac{1}{2}}\right\vert<1$, the last term in Eq.(\ref{MBexp}) is zero because of the occurence of $\Gamma(-2n)$ in the denominator. We emphasize that for special combinations of the parameters such as $a=b$ or $a=b+\frac{1}{2}$, that at first sight one might believe to be able to compensate the diverging behaviour of $\Gamma(-2n)$ due to $\Gamma(a-b-m)$ or $\Gamma\left(a-b-\frac{1}{2}-m-n\right)$ respectively in the numerator, Eq.(\ref{MBexp}) can in fact not be used. Indeed, for such special situations, the computation of the Mellin-Barnes representation  would have led to other formulas. These special cases are therefore tacitly discarded in the following. Note also that the special case where $a=2b$ is forbidden by definition, since it would violate the statement regarding the contour in Eq.(\ref{3F2MB}).

\begin{figure}[hbtp]
\centering
\includegraphics[scale=0.35]{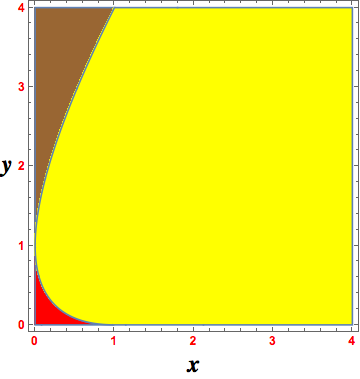}
\includegraphics[scale=0.35]{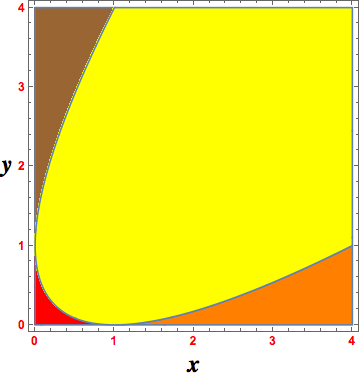}
\caption{Convergence regions $R_1$ (left) and $R_2$(right), in yellow.}
\label{Fig3}
\end{figure}

Using these results in Eq.(\ref{3F2avantMB}), we get, for Eq.(\ref{firstAC}):
\begin{align}
F_4(a,b;c,a-b+1;x,y)=M(a,b,c;x,y),
	\label{F4_AC2}
\end{align}
where  
\begin{align}
&M\left(a,b,c,x,y\right)\doteq(1-y)^{-a}\frac{\Gamma(a-b+1)\Gamma(c)}{\Gamma(a)\Gamma(b)}\left[\left(\frac{(1+y)^2}{(1-y)^2}\right)^{-\frac{a}{2}}\frac{2^{a-2b}}{\sqrt{\pi}}\frac{\Gamma\left(-b+\frac{1}{2}\right)\Gamma\left(\frac{1}{2}+b\right)}{\Gamma(a-2b+1)\Gamma(2b-a)}\right.\nonumber\\
&\times N\left(a,b,2b-a;b+\frac{1}{2},c;\frac{x}{2(1-y)}\left(\frac{(1-y)^2}{(1+y)^2}\right)^\frac{1}{2},\frac{(1-y)^2}{4(1+y)^2}\right)\nonumber\\
&+\left(\frac{(1+y)^2}{(1-y)^2}\right)^{-\frac{a}{2}+b-\frac{1}{2}}2^{a-1}\sqrt{\pi}\left\{-\left(\frac{2x}{y-1}\left(\frac{(1+y)^2}{(1-y)^2}\right)^{\frac{1}{2}}\right)^{-b}\frac{1}{\Gamma(c-b)\Gamma(1+b-c)}\right.\nonumber
\end{align}
\begin{align}
&\times N\left(a-b+1,1+b-c,b;\frac{3}{2},a-b+1;\frac{1-y}{2x}\left(\frac{(1-y)^2}{(1+y)^2}\right)^\frac{1}{2},\frac{(1-y)^2}{4(1+y)^2}\right)\nonumber\\
&+\left(\frac{2x}{y-1}\left(\frac{(1+y)^2}{(1-y)^2}\right)^{\frac{1}{2}}\right)^{\frac{1}{2}-b}\frac{1}{\Gamma\left(\frac{1}{2}+b-c\right)\Gamma\left(\frac{1}{2}-b+c\right)}\nonumber
\end{align}
\begin{align}
&\left.\left.\times O\left(a-b+\frac{1}{2},b-c+\frac{1}{2},b-\frac{1}{2};a-b+\frac{1}{2},\frac{1}{2};\frac{1-y}{2x}\left(\frac{(1-y)^2}{(1+y)^2}\right)^{\frac{1}{2}},\frac{x(1-y)}{2(1+y)^2}\left(\frac{(1+y)^2}{(1-y)^2}\right)^\frac{1}{2}\right)\right\}\right],\nonumber\\
	\label{M}
\end{align}
where we have defined
\begin{align}
N\left(\alpha,\beta,\gamma;\delta,\lambda;X,Y\right)\doteq\sum_{m,n=0}^{\infty}\frac{X^m}{m!}\frac{Y^n}{n!}\frac{\Gamma(\alpha+m+2n)\Gamma(\beta+m)\Gamma(\gamma+m)}{\Gamma(\delta+m+n)\Gamma(\lambda+m)}
	\label{N}
\end{align}
and
\begin{align}
O\left(\alpha,\beta,\gamma;\delta,\lambda;X,Y\right)\doteq\sum_{m,n=0}^{\infty}\frac{X^m}{m!}\frac{Y^n}{n!}\frac{\Gamma(\alpha+m+n)\Gamma(\beta+m-n)\Gamma(\gamma+m-n)}{\Gamma(\delta+m-n)\Gamma(\lambda+m-n)}.
	\label{O}
\end{align}
As shown before, the $N$ series converges like the Horn $H_3$ series. However, since the second $N$ series in the r.h.s of Eq.(\ref{M}) does not have the same arguments as the first, its convergence region 
\begin{align}
	R_1\doteq\left\vert \frac{(1-y)^2}{4(1+y)^2}\right\vert <\frac{1}{4} \wedge \left\vert \frac{1-y}{2x}\left(\frac{(1-y)^2}{(1+y)^2}\right)^{\frac{1}{2}}\right\vert <\frac{1}{2}+\frac{1}{2}\sqrt{1-4\left\vert \frac{(1-y)^2}{4(1+y)^2}\right\vert}
	\label{r1}
\end{align}
is not exactly the same as $R_0$, depicted in Fig. \ref{Fig2} (see the yellow region left figure in Fig. \ref{Fig3}). 

\begin{figure}[hbtp]
\centering
\includegraphics[scale=0.35]{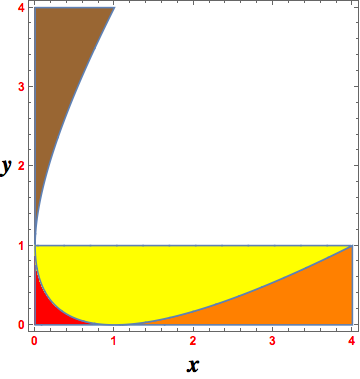}
\includegraphics[scale=0.35]{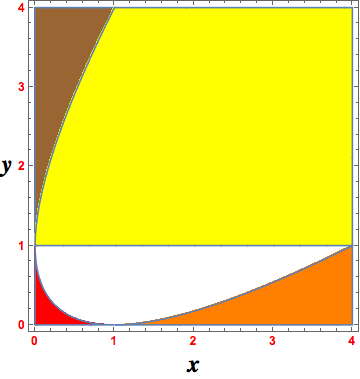}
\caption{Convergence regions $\tilde R$ (left) and $\tilde R'$ (right), in yellow.}
\label{Fig4}
\end{figure}

It is clear that the $O$ series in Eq.(\ref{O}) converges in the same way as the Appell $F_2$ does. The corresponding region of convergence
\begin{align}
	R_2\doteq\left\vert \frac{1-y}{2x}\left(\frac{(1-y)^2}{(1+y)^2}\right)^{\frac{1}{2}}\right\vert+ \left\vert \frac{x(1-y)}{2(1+y)^2}\left(\frac{(1+y)^2}{(1-y)^2}\right)^{\frac{1}{2}}\right\vert <1
	\label{r2}
\end{align}
is shown at the right side of Fig. \ref{Fig3}. The intersection of these two regions with the convergence region of Fig. \ref{Fig2} is therefore nothing but $R_2$ and fully covers the white region. However, it should be kept in mind that the quadratic formula Eq.(\ref{quadratic1}) is valid only for $\vert y\vert<1$ (the constraint coming from the convergence of the first two sums in the r.h.s of Eq.(\ref{MBexp}) can be ignored, since it is included in $R_2$). Therefore, although as we have just seen the series involved in $M$ do have a common convergence region in the $\vert y\vert>1$ zone of the $(x,y)$-plane, the region of validity of our new analytic continuation formula is restricted to\footnote{See the appendix for an illustration in a simple case.}
\begin{align}
\tilde R\doteq R_2\wedge\vert y\vert<1,
\end{align}
which is shown in the left pannel of Fig. \ref{Fig4}.

The first obvious idea to get an analytic continuation of $F_4$ valid in the region
\begin{align}
\tilde R'\doteq R_2\wedge\vert y\vert>1
\end{align}
is to follow the same derivation that led to Eq.(\ref{F4_AC2}), adding one extra step in the very beginning, by using the analytic continuation given in Eq.(\ref{F4_AC1}). Indeed, the latter relation has the nice property of preserving the specific pattern $d=a-b+1$ of the parameters of $F_4$, on which we have based all our analysis. We then obtain
\begin{align}
F_4(a,b;c,a-b+1;x,y)=&\frac{\Gamma(a-b+1)\Gamma(b-a)}{\Gamma(1-b)\Gamma(b)}(-y)^{-a}M\left(a,b,c;\frac{x}{y},\frac{1}{y}\right)\nonumber\\
+&\frac{\Gamma(a-b+1)\Gamma(a-b)}{\Gamma(a-2b+1)\Gamma(a)}(-y)^{-b}M\left(2b-a,b,c;\frac{x}{y},\frac{1}{y}\right).
	\label{F4_AC4}
\end{align}
Now, since the convergence regions $R_0, R_1$ and $R_2$ are symmetric under the simultaneous exchange of $x\leftrightarrow \frac{x}{y}$ and $y\leftrightarrow \frac{1}{y}$, nothing changes concerning them, except the constraint coming from the quadratic transformation formula which now becomes $\vert y\vert>1$. This is why the convergence region of the r.h.s of Eq.(\ref{F4_AC4}) is $\tilde R'$, depicted in the right side of Fig. \ref{Fig4}.

However, for real values of $x$ and $y$, Eq.(\ref{F4_AC4}) as it stands does not provide the right behavior on its branch cuts.
In the spirit of what is done at several instances in \cite{BS} for the Gauss hypergeometric function, one can solve this problem by the replacement
\begin{align}
\left(\frac{2x}{y-1}\left(\frac{(1+y)^2}{(1-y)^2}\right)^{\frac{1}{2}}\right)^{-b}&\longrightarrow \left(\frac{y-1}{2x}\left(\frac{(1-y)^2}{(1+y)^2}\right)^{\frac{1}{2}}\right)^{b}\nonumber\\
\left(\frac{2x}{y-1}\left(\frac{(1+y)^2}{(1-y)^2}\right)^{\frac{1}{2}}\right)^{\frac{1}{2}-b}&\longrightarrow \left(\frac{y-1}{2x}\left(\frac{(1-y)^2}{(1+y)^2}\right)^{\frac{1}{2}}\right)^{-\frac{1}{2}+b}
\label{replacement}
\end{align}
in the expression (\ref{M}) of the $M$ functions of the rhs of Eq.(\ref{F4_AC4}).

Note that this flip naturally comes by replacing eqs. (\ref{3F2MB}) by \cite{Erdelyi}
\begin{align}
    {}_pF_q & \left( \begin{array}{c}
a_1,...,a_p \\
b_1,...,b_q \\
\end{array} \bigg| z \right)
= \frac{\prod_{j=1}^q \Gamma(b_j)}{\prod_{j=1}^p \Gamma(a_j)}
G_{p,1}^{q+1,p} \left( -\frac{1}{z} \bigg|
\begin{array}{c}
1,b_1,...,b_q \\
a_1,..., a_p \\
\end{array} \right),
\end{align}
where $G^{m,n}_{p,q}$ is the Meijer-$G$ function. Indeed, applying the expansion 
\begin{align}
G_{p,q}^{m,n} & \left( z \bigg|
\begin{array}{c}
 a_1, ... , a_p \\
 b_1, ... , b_q \\
\end{array}
\right) = \sum_{h=1}^{m} \frac{\prod_{j=1}^{'m} \Gamma(b_j-b_h)
\prod_{j=1}^{n} \Gamma(1+b_h-a_j)}{ \prod_{j=m+1}^{q}
\Gamma(1+b_h-b_j) \prod_{j=n+1}^{p} \Gamma(a_j-b_h)} z^{b_h} \nonumber
\\
& \times {}_p F_{q-1} \left( \begin{array}{c}
1+b_h-a_1, ... , 1+b_h-a_p \\
1+b_h-b_1, ... , * , ... , 1+b_h-b_q  \\
\end{array} \bigg| (-1)^{p-m-n} z \right)
\end{align}
which is valid for $\vert z\vert<1$ and $p\leq q$ and where the prime in $\prod'$ indicates the omission of $\Gamma(b_r-b_r)$ and the asterisk in the $_qF_{p-1}$ the omission of the parameter $1+b_r-b_r$ \cite{Erdelyi}, one then obtains a similar formula, with $M$ replaced by $\tilde M$:
\begin{align}
F_4(a,b;c,a-b+1;x,y)=&\frac{\Gamma(a-b+1)\Gamma(b-a)}{\Gamma(1-b)\Gamma(b)}(-y)^{-a}\tilde M\left(a,b,c;\frac{x}{y},\frac{1}{y}\right)\nonumber\\
+&\frac{\Gamma(a-b+1)\Gamma(a-b)}{\Gamma(a-2b+1)\Gamma(a)}(-y)^{-b}\tilde M\left(2b-a,b,c;\frac{x}{y},\frac{1}{y}\right)
	\label{F4_AC3}
\end{align}
where the only difference between $M$ and $\tilde M$ comes from the replacements (\ref{replacement}) in Eq.(\ref{M}). The convergence region of Eq.(\ref{F4_AC3}) is the same as the one of Eq.(\ref{F4_AC4}).

The two new analytic continuations of $F_4$ obtained in Eqs.(\ref{F4_AC2}) and (\ref{F4_AC3}) then allow us to cover the white zone of Fig. \ref{Fig1}. As has been shown in \cite{Ananthanarayan:2019icl}, one can use these expressions to derive new series representations of the two-loop sunset diagram with four mass scales. As a last remark, we stress that alternative analytic continuations can be obtained, since other quadratic transformation formulas of ${}_2F_1$ do exist. This is the subject of the next section.

\section{Alternative analytic continuation formulas\label{other}}

In lieu of Eq.(\ref{quadratic1}) one can use the alternative quadratic transformation formulas \cite{A&S}, \cite{Erdelyi}
\begin{align}
{}_{2}F_{1}(\alpha,\beta;\alpha-\beta+1;y) & = (1- \sqrt{y})^{-2\alpha} {}_{2}F_{1}\left(\alpha, \alpha-\beta+ \frac{1}{2}; 2\alpha-2\beta+1;\frac{- 4 \sqrt{y}}{(1- \sqrt{y})^2}\right)
\label{quadratic2}\\
{}_{2}F_{1}(\alpha,\beta;\alpha-\beta+1;y) & = (1+ \sqrt{y})^{-2\alpha} {}_{2}F_{1}\left(\alpha, \alpha-\beta+ \frac{1}{2}; 2\alpha-2\beta+1;\frac{4 \sqrt{y}}{(1+ \sqrt{y})^2}\right)
\label{quadratic3}\\
{}_{2}F_{1}(\alpha,\beta;\alpha-\beta+1;y)& = (1+y)^{-\alpha} {}_{2}F_{1}\left(\frac{\alpha}{2},  \frac{\alpha}{2}+\frac{1}{2}; \alpha-\beta+1;\frac{4 y}{(1+y)^2}\right)
\label{quadratic4}\\
    {}_2F_1 \left(\alpha,\beta;\alpha-\beta+1;y\right)
 & = (1+y) (1-y)^{-\alpha-1} {}_2F_1  \left( 
\frac{1}{2}+\frac{\alpha}{2}, 1+\frac{\alpha}{2}-\beta ;
\alpha-\beta+1 ; \frac{-4y}{(1-y)^2} \right)
\label{quadratic5}
\end{align}
and
\begin{align}
    {}_2F_1 \left(\alpha,\beta;\alpha-\beta+1;y\right)
  = (1-y)^{1-2\beta} (1+y)^{2\beta-\alpha-1} {}_2F_1  \left( 
\frac{1}{2}+\frac{\alpha}{2}-\beta, 1+\frac{\alpha}{2}-\beta ;
\alpha-\beta+1 ; \frac{4y}{(1+y)^2} \right)
\label{quadratic6}
\end{align}
to see if they can give other relevant results. As an example, let us consider the first quadratic transformation (\ref{quadratic2}) (we postpone the study of the other quadratic transformations to future work).

Following the same steps of derivation as in Section \ref{AppellF4AC}, but using Eq.(\ref{quadratic2}) as starting point, we find at the end of the procedure some other analytic continuations of $F_4$ which do have the same final convergence regions as those shown in Fig. \ref{Fig4}, although the final expressions do not coincide with those described in Section \ref{AppellF4AC}. Let us give here the final expression in explicit form for the $\tilde R$ region:
\begin{align}
F_4(a,b;c,a-b+1;x,y)=M'(a,b,c;x,y)
	\label{F4_AC5}
\end{align}
with
\begin{align}
M'(a,b,c;x,y)&=(1+\sqrt{y})^{-2 a} \frac{2^{2a-2b}}{\sqrt{\pi}}\frac{\Gamma( a - b +1) \Gamma\left(-b+\frac{1}{2}\right)}{\Gamma(a-2b+1)}   \nonumber \\
&\hspace{1cm}\times F^{1:2;1}_{1:1;0} \bigg[ \begin{array}{c}
		a:b,2 b -a; a-b+\frac{1}{2}  \\
		b+\frac{1}{2}: c,\ - \\
	\end{array}	\bigg| \frac{x}{(1+\sqrt{y})^2} , \frac{(1-\sqrt{y})^2}{(1+\sqrt{y})^2} \bigg]   \nonumber\\
	&-\left((1-\sqrt{y})^2\right)^{\frac{1}{2}} \left((1+\sqrt{y})^2\right)^{-a+b-\frac{1}{2}} 2^{2a-2b+1}
\frac{\Gamma(a-b+1)\Gamma(c)}
{\Gamma(a) \Gamma(c-b)} (-x)^{-b} \nonumber \\
&\hspace{1cm}\times \left. F^{1:2;1}_{1:1;0} \bigg[ \begin{array}{c}
		1-b+a:b,1-c+b ; a-b+\frac{1}{2}  \\
		\frac{3}{2}:1-b+a: \ - \\
	\end{array}	\bigg| \frac{(1-\sqrt{y})^2}{x} , \frac{(1-\sqrt{y})^2}{(1+\sqrt{y})^2} \bigg]  \right) \nonumber\\
&+  \frac{((1+\sqrt{y})^2)^{-a+b-\frac{1}{2}} 2^{2 a -2 b}\sqrt{\pi}\Gamma(a-b+1) \Gamma(c) }{\Gamma(a)
 \Gamma(b) \Gamma\left(c-b+\frac{1}{2}\right)\Gamma\left(\frac{1}{2}+a-b\right)\Gamma\left(\frac{1}{2}+b-c\right)}(-x)^{-b+\frac{1}{2}}\nonumber\\
  &\times  P\left(\frac{1}{2}+a-b,\frac{1}{2}+a-b,b-\frac{1}{2},\frac{1}{2}+b-c;\frac{1}{2}+a-b,\frac{1}{2};\frac{(1-\sqrt{y})^2}{x},\frac{x}{(1+\sqrt{y})^2}\right)
 \label{M'}
 \end{align}
where the $F^{1:2;1}_{1:1;0}$ series is a generalized Kamp\'e de F\'eriet series \cite{Srivastava} and
\begin{align}
P\left(\alpha,\beta,\gamma,\delta;\lambda,\mu;X,Y\right)\doteq\sum_{m,n=0}^{\infty}\frac{X^m}{m!}\frac{Y^n}{n!}\frac{\Gamma(\alpha+m)\Gamma(\beta+n)\Gamma(\gamma+m-n)\Gamma(\delta+m-n)}{\Gamma(\lambda+m-n)\Gamma(\mu+m-n)}.
	\label{P}
\end{align}
We note here that we have performed some numerical cross-checks between the expressions given in Eq.(\ref{F4_AC2}) and those in Eq.(\ref{F4_AC5}) and we obtained a perfect agreement.

\section{Conclusion \label{concl}}

The Appell series, $F_i$ for $i=1,2,3,4$, are a set of double series that were historically introduced to extend the Gauss hypergeometric series to two variables. They are special cases of Kamp\'e de F\'eriet series. These objects are widely used in both theoretical and applied mathematics, engineering and physics. The standard definition of these functions is in the form of a double infinite series, which converges for a specified range of values of the variables. Therefore, to obtain series representations of the $F_i$ beyond the range of validity of their standard series definitions, analytic continuation is necessary. 
In the case of $F_2, F_3$ and $F_4$, a direct residue evaluation of its MB integral representation gives rise to series representations that allow only a partial extension of this range of validity (for $F_1$ and some other Kamp\'e de F\'eriet series, this is not the case \cite{Ananthanarayan:2020acj}). An alternate approach, involving the analytic continuation of one of the two sums constituting the $F_4$ series, has yielded series valid in the region which lies out of the convergence regions of the basic series definition of $F_4$ and of its analytic continuations that can be obtained from its MB representation \cite{Exton}. In this work, for the special case when $d=a-b+1$, we follow a similar approach to derive two analytic continuation formulae for $F_4(a,b;c,d;x,y)$ that cover the $x>0, y>0$ domain, useful for phenomenological applications in quantum field theory, as well as other regions with negative and/or complex values of $x$ and $y$. 

Our analytic continuation technique consists of rewriting one the two sums that makes up the $F_4$ series as a $_2F_1$ hypergeometric series and analytically continuing the latter after applying some of its quadratic transformations. These quadratic formulae are valid only for small values of the argument, which therefore requires application of additional relations for the cases where the $y$ variable of the $F_4$ function is greater than 1 in absolute value, and which thus leads us to two distinct expressions for the analytic continuation of $F_4$ (for small, and for large, values of the $y$ variable). Since there exist several quadratic relations for the $_2F_1$ hypergeometric function, use of each leads to different analytic continuations of $F_4$. Some of these were used to test and verify the results of this paper.

The method used in this paper is important due to its wide applicability, also in areas beyond the theoretical particle physics which was our primary motivation. Partly this is due to the fact that special functions such as the hypergeometric, Kamp\'e de F\'eriet and their generalisations are of widespread use. 
In the quantum field theory context, it has been shown in \cite{Ananthanarayan:2019icl} that Eq.(\ref{F4_AC2}) of the present paper is well suited for an application to the study of the two-loop sunset diagram with four mass scales, in order to derive new triple series representations of the latter valid, among others, in chiral perturbation theory and in some sectors of the minimal supersymmetric standard model. 
It is also possible to apply our analytic continuation procedure to the expressions of the three mass scales sunset diagrams given in \cite{Ananthanarayan:2017qmx,Ananthanarayan:2018irl}, as will be shown in \cite{AFGkaon}.

\vspace{1cm}

{\bf Acknowledgements}
S.F. thanks David Greynat for discussions on some topics related to the content of this paper.
B.A. acknowledges partial support from the Mysore Sales International Limited Chair of the Division of Physical and Mathematical
Sciences, Indian Institute of Science during the course of this work.  S.G. thanks Ulf-G. Meissner for supporting the research through grants. S.G. and A.H. thank Institut de Physique Nucléaire d'Orsay, Université Paris-Sud
and  S.F. thanks Centre for High Energy Physics, Indian Institute of Science of Bangalore for their hospitality during
the course of this work. A. H. thanks CEFIPRA for awarding him of the Raman-Charpak fellowship for his internship in Centre for High Energy Physics, Indian Institute of Science of Bangalore, during summer 2019.

\section*{Appendix \label{Appendix}}

In this Appendix we illustrate, in a simple example, some subtelties about the range of validity of transformation (or analytic continuation) formulas and the convergence properties of the series involved in the latter.
 
\begin{figure}[hbtp]
\centering
\includegraphics[scale=0.35]{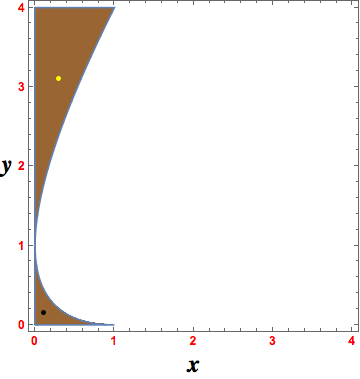}
\caption{Convergence region $R$. In black, the point $(x=0.1,y=0.15)$, in yellow, the point $(x=0.3,y=3.1)$.}
\label{Fig6}
\end{figure}

Let us consider Eq.(97) p. 303 of \cite{Srivastava}:

\begin{equation}
\label{example_Sri}
F_4\left(a,a-b+\frac{1}{2},c,b+\frac{1}{2};x,y\right)=(1+\sqrt{y})^{-2a}F_2\left(a,a-b+\frac{1}{2},b;c,2b;\frac{x}{(1+\sqrt{y})^2},\frac{4\sqrt{y}}{(1+\sqrt{y})^2}\right)
\end{equation}
This result between Appell functions has been obtained by applying the $+$ sign version of Eq.(\ref{quadratic2}) to the LHS of Eq.(\ref{example_Sri}). The convergence region of the $F_2$ series in the RHS of Eq.(\ref{example_Sri}) is simply given by 
\begin{equation}
R=\left\vert\frac{x}{(1+\sqrt{y})^2}\right\vert+\left\vert\frac{4\sqrt{y}}{(1+\sqrt{y})^2}\right\vert<1
\end{equation}
and is depicted for positive values of $x$ and $y$ in Figure \ref{Fig6}.

In Fig \ref{Fig6}, we see that the RHS of Eq.(\ref{example_Sri}) is converging in the usual convergence region of the Appell $F_4$ function (the red region in Fig \ref{Fig1}), but also in a region where $y>1$ (the upper brown region of Fig \ref{Fig1}) which, in fact, is nothing but the convergence region of Eq.(\ref{F4_AC1}).
This does not mean, however, that Eq.(\ref{example_Sri}) is an analytic continuation valid in the latter and, as we will see, it is not, the reason being that Eq.(\ref{quadratic2}) is not valid for $y>1$.

To illustrate this, let us perform the numerical evaluation of the LHS and the RHS of Eq.(\ref{example_Sri}) for $a=1.7$, $b=1.7$ and $c=1.1$, in both the lower and upper regions of Fig. \ref{Fig6} using the built-in Appell $F_4$ and $F_2$ functions of $\mathsf{Maple}$ (we have cross-checked $\mathsf{Maple}$'s results with $\mathsf{Mathematica}$ by evaluating the corresponding truncated Appell $F_4$ series of Eq.(\ref{F4}) in the lower region, and of Eq.(\ref{F4_AC1}) in the upper one, against the common (truncated) series representation of $F_2$).

In the lower (and common) region of convergence of both the $F_4$ and $F_2$ series, choosing for instance $x=0.1$ and $y=0.15$ (the black point in Fig. \ref{Fig6}), one obtains the following results:
\begin{equation}
\label{LHS}
F_4\left(1.7,\frac{1}{2},1.1,2.2;0.1,0.15\right)\simeq 1.18614
\end{equation}
and
\begin{equation}
\label{RHS}
(1+\sqrt{0.15})^{-3.4}F_2\left(1.7,\frac{1}{2},1.7;1.1,3.4;\frac{0.1}{(1+\sqrt{0.15})^2},\frac{4\sqrt{0.15}}{(1+\sqrt{0.15})^2}\right)\simeq 1.18614
\end{equation}
which, as expected, shows a perfect agreement between the RHS and the LHS of Eq.(\ref{example_Sri}). 

In the upper region, choosing for instance $x=0.3$ and $y=3.1$ (the yellow point in Fig. \ref{Fig6}), one obtains\footnote{The point $(x=0.3,y=3.1)$ does not belong to the convergence region of the $F_4$ series, therefore the result given in Eq.(\ref{LHS2}) by the built-in $F_4$ function of $\mathsf{Maple}$ has been obtained by analytic continuation of the $F_4$ series.}:
\begin{equation}
\label{LHS2}
F_4\left(1.7,\frac{1}{2},1.1,2.2;0.3,3.1\right)\simeq 0.19925- 0.88214i
\end{equation}
and
\begin{equation}
\label{RHS2}
(1+\sqrt{3.1})^{-3.4}F_2\left(1.7,\frac{1}{2},1.7;1.1,3.4;\frac{0.3}{(1+\sqrt{3.1})^2},\frac{4\sqrt{3.1}}{(1+\sqrt{3.1})^2}\right)\simeq 0.19925
\end{equation}
We see that the imaginary part of the LHS, in Eq.(\ref{LHS2}), has not been reproduced by the RHS, in Eq.(\ref{RHS2}).
In passing, it is not surprising that Eq.(\ref{LHS2}) develops an imaginary part because the Appell $F_4$ series has branch cuts on both the $x$ and $y$ real axis, running from $1$ to $+\infty$. 

As said above, to check the result of Eq.(\ref{LHS2}) with $\mathsf{Mathematica}$ using truncated series expressions, one can use the analytic continuation given in Eq.(\ref{F4_AC1}), which is valid for $x=0.3$ and $y=3.1$ and where, once applied to our case of study, one can notice that overall contributions of the type $(-3.1)^{-1.7}$ and $(-3.1)^{-\frac{1}{2}}$ appear. Since these objects correspond fractional powers evaluated on their branch cut, this explains the imaginary contribution. On the contrary, in Eq.(\ref{RHS2}) the overall fractional power factor is real. 

The conclusion is that although the Appell $F_2$ series is converging in the upper region of Fig. \ref{Fig6} where $y>1$, one cannot use Eq.(\ref{example_Sri}) as an analytic continuation of the corresponding Appell $F_4$ series in this region.

\end{document}